**An Empirical Analysis of how Internet Access Influences Public Opinion towards Undocumented Immigrants and Unaccompanied Children**

**Muhammad _Hassan_ Bin Afzal**
**Department of Political Science**
**Kent State University**
**Kent, Ohio, 44242**
**mafzal@kent.edu**
**https://orcid.org/0000-0001-8192-0885**

**<u>Corresponding Author:</u>**
**Muhammad _Hassan_ Bin Afzal**
**Department of Political Science**
**Kent State University**
**Kent, Ohio, 44242**
**mafzal@kent.edu**
**https://orcid.org/0000-0001-8192-0885**

**Final Word Count: 5000**

# An Empirical Analysis of how Internet Access Influences Public Opinion towards Undocumented Immigrants and Unaccompanied Children

## Muhammad *Hassan* Bin Afzal

***Abstract*—** This research adds to the expanding field of data-driven analysis, scientific modeling, and forecasting on the impact of having access to the Internet and IoT on the general US population regarding immigrants and immigration policies. More specifically, this research focuses on the public opinion of undocumented immigrants in the United States and having access to the Internet in their local settings. The term "Undocumented Immigrants" refers to those who live in the United States without legal papers/documents/visas. Undocumented immigrants may have come into the country unlawfully or with valid documentation, but their legal status has expired. Using the 2020 American National Election Studies (ANES) time series dataset, I investigated the relationship between internet access (A2I) and public perception of undocumented immigrants. According to my research and analysis, increasing internet access among non-Hispanic whites with at least a bachelor's degree with an annual household income of less than 99K is more likely to oppose the deportation of undocumented immigrants and separating unaccompanied children from their families in borderland areas. The individuals with substantial Republican political ideology exhibit significantly lower opposing effects in deporting undocumented immigrants and/or separating unaccompanied children from their families. The evidence from multiple statistical models is resilient to a variety of factors. The findings show that increased internet access may improve undocumented immigrants' social integration and acceptability and that it may be especially beneficial during health emergencies to make them feel safe, included, and supported in their local settings.

## I. INTRODUCTION

The ongoing COVID-19 epidemic has aggravated the mobility problem and hampered access to essential services for immigrants globally due to a lack of adequate legal documents and perceived societal concern [1], [2]. Pre-existing socio-economic inequalities for disadvantaged individuals in Latin American nations (LAC) increased during COVID-19, driving a higher migration rate to other countries seeking safety, healthcare, and better life prospects. Unfortunately, certain politicians' and lawmakers' narratives regarding COVID-19 have resulted in a more divided and unfavorable public view of immigrants, and a sense of "otherness" has grown critical. Several empirical investigative studies discovered that previous health crises resulted in substantial social and economic inequities and unequal access to healthcare and vaccinations for immigrants. Similarly, the continuing COVID-19 has a similar impact on underprivileged and vulnerable immigrant populations across the world. As a result, low-income and vulnerable Americans will bear a disproportionate share of the burden. Minority populated communities, particularly poverty-stricken immigrant communities, are already suffering from higher infection rates, severe lack of access to healthcare, lower vaccinations, and limited access to public socio-economic support [1], [3]–[7].

To that objective, the Democratic Party worked hard and fostered grassroots efforts to secure bipartisan support for the infrastructure package. President Biden highlighted and campaigned for the infrastructure bill throughout his presidential campaign. With a lengthy history in the United States Senate and politics, President Biden and the Democratic Party worked hard all spring and summer of 2021 to negotiate and find a middle ground to enact the ambitious infrastructure package with significant bipartisan support. [8]–[10]. The 2,702-page bipartisan bill includes only $550 billion in increased government spending. The $1.2 trillion total is generated from increased funding granted each year for highways and other public works projects. Roads and bridges will receive $110 billion in additional investment, railways will receive $66 billion, the energy system will receive $65 billion, enhanced high-speed Internet will receive $65 billion, and cybersecurity and environmental considerations will receive $47 billion. The $65 billion allocated for broadband includes funds to expand internet access in rural regions and low-income communities. Approximately $14 billion of the funds would be used to reduce the cost of internet access for low-income households.

One of the critical components of the infrastructure bill is improving and bettering internet access in rural and underserved U.S. areas to provide better access to education, healthcare, and the labor market. As a result, the primary goal of this research is to investigate and document the effect of Internet access in shaping public opinion of immigrants and immigration policy in the United States. The recently published American National Election Studies (ANES) 2020 time-series data set was utilized to capture the effect of access to the Internet and public perception towards undocumented immigrants. In successive models, a set of social demographic variables (age, race, income, and education level), nativeness, and political ideology were tested [11].

The results find that accessing the Internet reduces unfavorable feelings towards undocumented immigrants among the general public. One possible explanation is that having regular access to the Internet either in their home or office or both would provide more information and visual representation of the struggles the undocumented immigrants are facing on a daily basis and humanize the situation. The general population feels less unwelcome and opposing towards undocumented immigrants and becomes more aware of and empathetic towards their well-being. The following section explores the current condition of access to the Internet in the USA and associated political narratives and policy developments, incorporating the following applicable criteria.

## II. UNDERSERVED COMMUNITIES AND INTERNET ACCESS

The communities of color have consistently lacked access to the Internet. More particularly, the marginalized and underserved rural communities with lower income and inadequate access to education also severely lacked internet access. Black and Latino communities in the U.S. consistently lacked access to the Internet in the last decade, and it highlights the ongoing digital divide based on race, location, education, and income [4], [12]. Figure-1 outlines the disparity and unequal access to the Internet-based on race.

Furthermore, various government-sponsored access to internet programs in the USA facilitates internet connection to the lower-income houses at lower and affordable prices. Notably, during the ongoing COVID-19 pandemic, the U.S. government initiated various local, state, and federal level programs to facilitate internet access to the rural, underserved, and remote communities [13]–[18]. Still, the rural and underserved communities lack the proper access to the Internet, which causes various barriers in getting adequate access to healthcare, public support services, educational materials, accurate information, and timely public health guidance. According to the 2021 PEW Study, suburban and rural areas lack access to the Internet compared to urban areas [12], [19]. This study mainly focuses on having internet access to general populations and how that impacts their perception and attitude towards immigrants and well-being and various immigration policies.

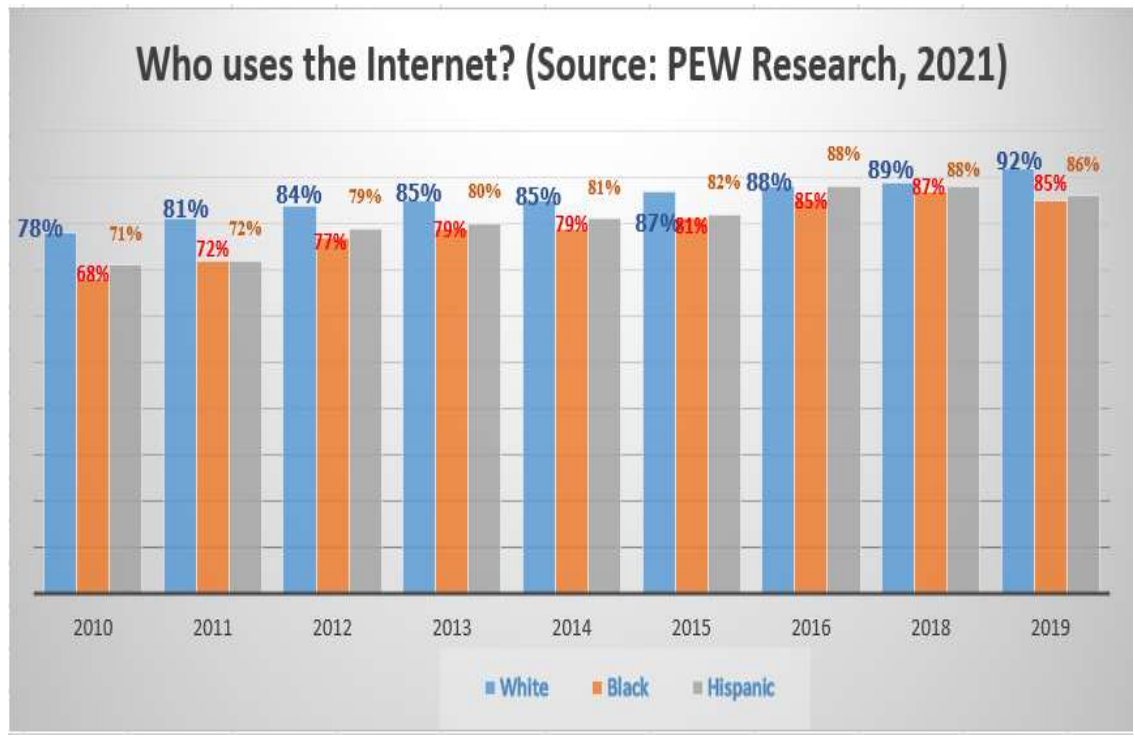

**Figure 1**: Internet usage (2010-2019)

Figure-1 explores the access to broadband Internet in the USA based on race[12]. Both black and Latino communities consistently lack equitable access to broadband Internet over the years. Unfortunately, internet usage has reduced from 2018 to 2019 among both Latino and Black communities by 2%. Various underlying factors cause the gaps in internet usage in Latino and black communities, such as political, religious, economic, social, technology, legal, and environmental. Lower-income families might perceive the associated cost of having broadband internet as a luxury item and decide against getting Internet at their homes and/or communities. The following section focuses on the study design and models for the empirical analyses.

### III. THE DESIGN OF THE STUDY

The data for this study were collected from the 2020 American National election studies (ANES) time-series dataset [11]. A group of social scientists and political researchers conduct surveys and collect data before and after U.S. elections. The ANES datasets are robust and functional to explore, analyze, and understand public perceptions about various public policies and social issues. I utilized four ordered logistic regression models for this study. My primary dependent variable for this study is the public sentiment (**pplSentim**). The general sentiment variable is the ordinal variable where O captures for the survey respondent do not oppose deporting undocumented immigrants from the country and separating accompanied children from their families in border areas. The value 1 captures for the survey respondent opposes either deporting the undocumented immigrants are separating accompanied children from their families in border areas. The value 2 frames for the survey respondent opposes both deporting undocumented immigrants and separating accompanied children from their families in border areas.

The primary independent variable for this study is access to the Internet (**A2I**), reach captures whether the survey respondent has access to the Internet either in their home or some other sources. The variable *A2I* is a dichotomous variable where 1 captures the server respondent has access to the Internet either in their home setting or office and vice versa. The second independent variable (*R_WNHispanic*) is also a binary variable where one captures that the survey respondent is white and vice versa. The third binary/dichotomous independent variable (*R_olderWAP*) captures that the survey respondent is a working-age professional aged 45 and 64 or not. The fourth binary/dichotomous independent variable (*R_mdincome*) captures whether the survey respondent's annual household income is below 99K or not. The fifth binary/dichotomous independent variable (*R_Bachedu*) captures whether the survey respondent earned at least a bachelor's degree or not. The sixth binary/dichotomous independent variable (***R_nativeness***) captures whether the survey respondent's at least one parent was born in the USA or not. The seventh binary/dichotomous independent variable (***R_republican***) captures whether the survey respondent's political ideology aligns with that U.S. republican party or not. Figure 2 outlines the summary statistics of all the variables used in all four ordered logistic regression models.

      ***Model 1*** only captures the effect of having access to the Internet, and public sentiment and perception towards undocumented immigrants add policies associated with separating accompanied children from their families in border areas. Similarly, ***Model 2*** includes social-demographic variables of the survey respondents such as race, age, annual income level, and education level. ***Model 3*** consists of the nativeness independent variable. Finally, ***Model 4*** has political ideology variable wheat nativeness social-demographic variables. The central research question is stated below.

    <u>**Research Question:**</u> How does access to the internet (A2I) shape non-Hispanic whites' sentiments about undocumented immigrants' well-being during health crises (COVID 19)?

    <u>**The Analytical Model:**</u> **Ln[prob(Public Sentiment)/ 1- prob(Public Sentiment)]$_{I,j,t}$**

$=$ Constant $(\alpha)+ \boldsymbol{\beta}$ (A2I)$_{I,j,t}$ + $\boldsymbol{\beta_1}$ (Age) $_{I,j,t}$ + $\boldsymbol{\beta_2}$(Race) $_{I,j,t}$, + $\boldsymbol{\beta_3}$(Education) $_{I,j,t}$, + $\boldsymbol{\beta_4}$(Annual Household Income) $_{I,j,t}$ + $\boldsymbol{\beta_5}$ (Political Ideology) $_{I,j,t}$ + $\boldsymbol{\beta_6}$ (Nativeness) $_{I,j,t}$

    In my models, (Public Sentiment)$_{I,j,t}$ is a categorical qualitative variable, the logarithm of the number of survey respondents (i) living in country j (the U.S.) in the year at (2020). The timeline for collecting 2020 ANES time series data started in August 2020 and continued until Nov. 03, 2020. The pre-election the post-election data collection process started right after Election Day and continued until the end of Dec. 31, 2020. The U.S. government officials began sharing information about the ongoing pandemic of 19 since the last week of February 2020.

That is why the 2020 time-series data is significant because both pre and post elections and the data collecting process started in the middle of the ongoing pandemic COVID-19. The survey participants were already exposed to the impacts of the ongoing health crisis and shared their opinions related to various public policies, social policies, and role minority communities. The harassment and misbehaving, medical coverage, race politics, border control, press credibility and disinformation, administrative legality, propaganda, political images, trade barriers, and tax policy variables have been included in the 2020 pre-election ANES questionnaire. Table 1 outlines the summary statistics of the variable used in the models.

TABLE 1.The Summary Statistics of the variables used in models

| Variable | No. of Obs | Mean | Std. Dev. | Min | Max |
|---|---|---|---|---|---|
| pplSentim | 6,359 | 1.186193 | 0.7269092 | 0 | 2 |
| A2I | 6,359 | 0.9175971 | 0.2749992 | 0 | 1 |
| R_WNHispanic | 6,359 | 0.7320333 | 0.442935 | 0 | 1 |
| R_olderWAP | 6,359 | 0.3352728 | 0.4721229 | 0 | 1 |
| R_mdincome | 6,359 | 0.4057242 | 0.4910703 | 0 | 1 |
| | | | | | |
| R_Bachedu | 6,359 | 0.260261 | 0.4388115 | 0 | 1 |
| R_nativeness | 6,359 | 0.0602296 | 0.2379304 | 0 | 1 |
| R_republican | 6,359 | 0.2028621 | 0.4021623 | 0 | 1 |

New topics for the 2020 census also include voting encounters, mentalities toward public health officials and institutions, anti-elitism, trust in scholars, environmental issues, gun regulation, prescription drugs, rural-urban identity, global trade, gender discrimination, and MeToo, non-binary military service, perception of overseas nations, group compassion and understanding, the relationship between social media, misleading information, and complacency. Furthermore, the 2020 Pew Research and the ANES datasets show that non-Hispanic white Americans have about 90% access to the Internet via mobile or home devices [11], [12]. The following section explores the findings and interpretations of the conclusions of the mathematical models.

### IV. THE STUDY FINDINGS

*Model 1* presents a simple bivariate model between the respondent's current access condition and their perception of undocumented immigrants. Notice the large OR value of 1.55. The 1.55 OR value indicates that when the respondent has access to the Internet either in their house and/or through their mobile phones, they are **55%** times more likely to oppose deporting the undocumented immigrants and/or separating children from their detained immigrant parents. Both the p-value and 95% confidence interval show the statistical significance of this model. _Model 2_ depicts that regardless of the social demographic variables, having access to the Internet, either home-internet or mobile phone results in less unfavorable feelings for undocumented immigrants and the immigration policies of unaccompanied children from their detained parents.

*Model 2* includes social-demographic variables such as race, age, yearly income, and degree of education achieved. The inclusion of social demographic variables in model 2 led to more welcoming and favorable attitudes from survey respondents regarding undocumented immigrants and the opposing sentiments towards the Trump administration's continued policy of removing unaccompanied children from their detained parents in 2020.

Table 2: The Comparative Statistical Output of all Models

| | | *Model 1* | *Model 2* | *Model 3* | *Model 4* |
|---|---|---|---|---|---|
| **pplSentim (Dependent Variable)** | | | | | |
| | | *b/se* | *b/se* | *b/se* | *b/se* |
| | | Odds Ratio | Odds Ratio | Odds Ratio | Odds Ratio |
| Variable (IV) | | | | | |
| **A2I** | Internet Access (Home or Mobile or both) or no access (1/0) | 1.555*** | 1.556*** | 1.555*** | 1.580*** |
| | | (-0.13) | (-0.13) | (-0.13) | (-0.14) |
| **R_WNHispanic** | Non-Hispanic White or not (1/0) | | 0.836*** | 0.840** | 1.123* |
| | | | (-0.04) | (-0.05) | (-0.06) |
| **R_olderWAP** | The Older working-age professional (45 yrs.+) or not (1/0) | | 0.788*** | 0.789*** | 0.804*** |
| | | | (-0.04) | (-0.04) | (-0.04) |
| **R_mdincome** | Annual HH Income < 99K or not (1/0) | | 0.905* | 0.906* | 0.951 |
| | | | (-0.04) | (-0.04) | (-0.05) |
| **R_Bachedu** | At least a bachelor's degree or not (1/0) | | 1.405*** | 1.404*** | 1.333*** |
| | | | -0.08 | -0.08 | (-0.07) |
| **R_nativeness** | At least One parent born in the US or not (1/0) | | | 1.098 | 1.038 |
| | | | | (-0.11) | (-0.11) |
| **R_republican** | Strong Republican Party Supporter or not (1/0) | | | | 0.172*** |
| | | | | | (-0.01) |
| | cut1 | | | | |
| | constant | 0.346*** | 0.290*** | 0.293*** | 0.219*** |
| | | (-0.03) | (-0.03) | (-0.03) | (-0.02) |
| | | | | | |
| | cut2 | | | | |
| | constant | 2.507*** | 2.141*** | 2.161*** | 2.053*** |
| | | -0.21 | -0.2 | -0.2 | -0.2 |
| | | | | | |
| | Number of obs | 6,359 | 6,359 | 6,359 | 6,359 |
| | Prob > chi2 | 0 | 0 | 0 | 0 |
| | LR chi2 | LR chi2(1) = 27.29 | LR chi2(5) = 103.56 | LR chi2(6) =104.43 | LR chi2(7) = 926.05 |
| | | | | | |
| | AIC | 13257 | 13189 | 13190 | 12371 |
| | BIC | 13277 | 13236 | 13244 | 12431 |
| | | | | | |
| | * p<0.05, ** p<0.01, *** p<0.001 | | | | |
| | The standard error (S.E.) is the number in parentheses. | | | | |

The survey respondent who has access to the Internet through their home and/or mobile phone is more likely to oppose deporting undocumented immigrants or separating children from their detained parents in 2020. The odds ratio of public sentiments for all four models is almost similar, approximately ranging from 1.55 ~ 1.58. The fourth or **complete model** shows that the non-Hispanic whites with at least a bachelor's degree and median household income oppose deporting undocumented immigrants and/or separating children from their detained parents.

Another interesting finding from **Model 4** is the proportional odds ratio of comparing strong Republican supporters to other political ideologies, given that the other variables in the model are held constant. According to the 4th model, strong republican supporters are less likely to oppose the deportation of undocumented immigrants and/or separating unaccompanied children from their detained parents. For strong Republicans, the odds of opposing the deportation of undocumented immigrants(U.I.) and separating unaccompanied children (UC) from their parents are (1- 0.17 = 83%) or 83% less likely to occur, holding all other variables constant and this statistical output is statistically significant at 0.001 level.

## V. RESULT ANALYSIS AND LIMITATIONS OF THE STUDY

**Model 4** shows that having access to the Internet, either in-home or mobile phone, increases opposing mentality regarding undocumented immigrants' deportation or separating children from their detained parents in border areas. Holding all else equal non-Hispanic whites (**R_WNHispanic**) with internet access are 12.3% likely to oppose either deportation of U.I. or separation of U.C., and it is statistically significant at 0.05%. This shows that exposure to media, the Internet, and daily news may facilitate the process of humanization and acceptance of underserved and marginalized U.I. and U.C., particularly during health crises (COVID-19). Holding all else equal working-age professionals aged 45 years or more (**R_olderWAP**) with internet access are 20% less likely to oppose both deportations of U.I. or separation of U.C., and it is statistically significant at 0.001%. Holding all else equal working-age professionals aged 45 years or more (**R_olderWAP**) with internet access are 20% less likely to oppose both deportations of U.I. or separation of U.C., and it is statistically significant at 0.001%. Holding all else equal participants with at least a bachelor's degree (**R_Bachedu**) with internet access are 33% more likely to oppose either deportation of U.I. or separation of U.C., and it is statistically significant at 0.001%. Holding all else equal participants with at least annual household income of <99K (**R_mdincome**) with internet access are 5% less likely to oppose either deportation of U.I. or separation of U.C. It is not statistically significant in Model 4. Figure 2 outlines the predicted outcome of having Internet and opposing views on deporting U.I. and separating U.C.

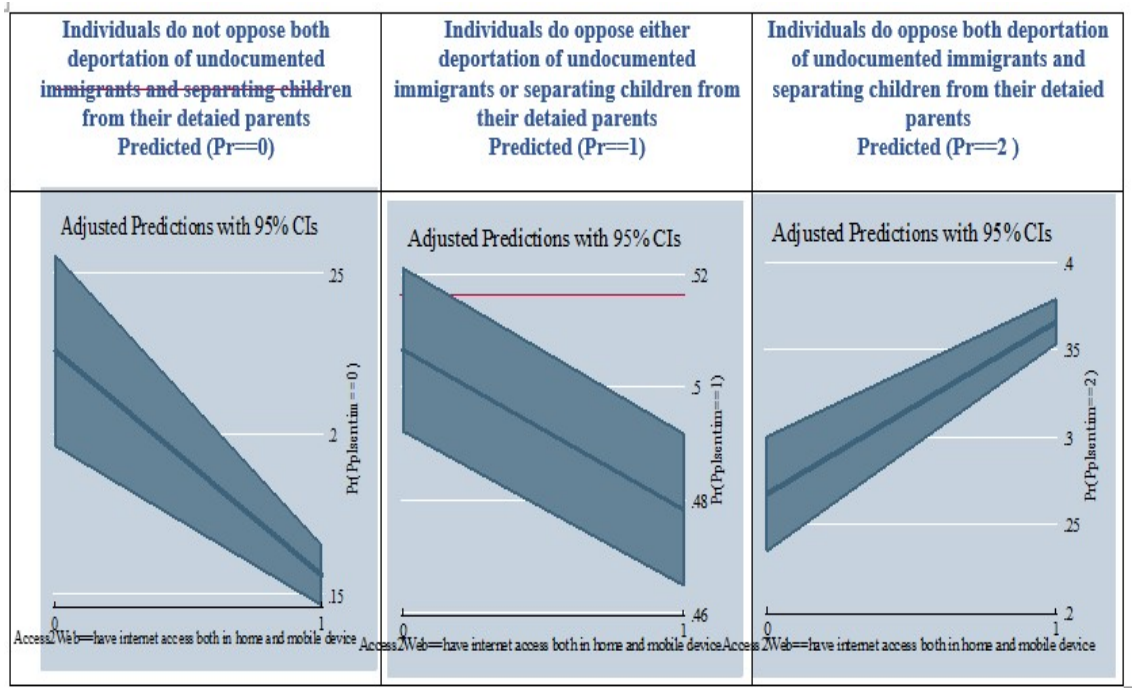

Fig 2: predicted probabilities of having access to the Internet

Figure-2 shows that having access to the Internet both in-home and mobile phones increase the predicted probability of opposing the deportation of U.I. and separating U.C. from their detained parents in border areas. Improved access to the Internet has a more profound and meaningful impact in creating peaceful and mutually accepting coexistence between immigrant and host communities. There are several limitations to this study. This study primarily focuses on the 2020 ANES time series data set, which is just a snapshot of the overall situation, context, and environment. It lacks the ability to capture the nuances add multi-faceted social-economic factors that could contribute to a more welcoming and accepting attitude towards undocumented immigrants and unaccompanied children. Regardless of the limitations, this study is a good foundational benchmark to assess and recognize the importance of having better Internet access for the general population. Situation and context-specific adequate data collection and more localized factors need to be considered and analyzed to understand the impact of the Internet and public attitude towards undocumented immigrants.

Access to the Internet and Internet-based services removes geographical and temporal restrictions, allowing people in rural areas to participate in socio-economic and sociopolitical activities that reach far beyond their physical location. Access to the Internet removes the practical constraints of regionally-based income sources, allowing small-town businesses to compete with big-city rivals. Moreover, the numerous potential benefits could be made available by helping high-speed Internet connection in rural areas of the entire country in the long run [20], [21].

This study finds that the effect of political ideology supersedes the impact of having high-speed Internet in their home settings or handheld devices. Even with an increased level of education and higher household annual income steal the political ideology impacts the survey respondents' impression and mindset towards undocumented immigrants and the policy of separating accompanied children in the U.S. The race of undocumented immigrants plays a significant role in shaping public perception and acceptance. Non-white Latin immigrants face harsher social reactions and stricter policies compared to white immigrants. Lower-income communities in developing countries face increasing barriers in accessing healthcare support [22], adequate sanitation facilities [23], and socio-economic support [24] during unexpected health crises or climate crises[25]. A 2021 Brookings study points out that non-Hispanic white undocumented immigrants receive better treatment and social acceptance than non-white Latino immigrants from developing Latin American countries [26].

The statistical models in this study found that non-Hispanic whites' educational attainment and steady annual household income combined with regular access to the Internet facilitates humanizing undocumented immigrants and accompanying children. The social acceptance of the immigrants among their host communities is not generational. This study also finds that the survey respondent's nativeness does not impact the perception of immigrants and immigration policies in a statistically significant manner. Regardless of having high-speed Internet, bachelor's education, and steady annual household income, an individual's political ideology plays a superseding role in shaping and forming the perception and attitude towards immigrants, the well-being of unaccompanied children, and immigration policies. Political parties play a significant role in reshaping and recreating public perception towards immigrants and immigration policies to attain inclusive socio-economic growth and social cohesion. Therefore the recently U.S. Senate-passed infrastructure bill could significantly enhance social cohesion between immigrant and host communities.

The narrative and rhetoric used by the lawmakers and politicians influence individuals' emotional perception and attitude towards immigrants, undocumented children, and immigration policies. Particularly during health crises the undocumented immigrants and unaccompanied children are some of the most vulnerable individuals. Due to perceived social fear, legal fear, and political rhetoric, these marginalized and underserved communities always fear reaching out and asking for help to navigate this unpredicted and challenging situation [3], [26]–[33]. Better access to the Internet and more sustainable infrastructure to disseminate high-speed Internet in remote and rural areas would improve public perception toward immigrants' immigration policies and the well-being of unaccompanied children, particularly during a health crisis. A potential explanation could be that access to high-speed Internet increases the exposure to the hardship and challenges the undocumented immigrants and unaccompanied children face during health crises.

Access to news media video clips an updated report humanizes the obstacles and barriers the undocumented immigrants and unaccompanied children regularly experience during a health crisis. Other countries worldwide could device situation-specific approaches that enhance access to the Internet, which ultimately influences social cohesion and social integration among immigrant communities and host communities. Finally, the rhetoric and narrative used by the lawmakers, politicians, political parties have a substantial lasting impact in shaping public attitude and sentiments towards immigrants', immigration policies, and well-being of the minority and underserved community members. Political parties and politicians need to be more mindful and observant about propagating information and ideas through their platforms so that these messages and ideologies do not harm, exclude, and isolate minorities and underserved communities. Better and inclusive messages based on scientific information and analysis disseminated through the platform of political parties and politicians improve and enhance community integration, collaborative community development, and sustainable socio-economic growth in the long run.